\documentstyle[12pt,epsf]{article}
\def\fo{\hbox{{1}\kern-.25em\hbox{l}}}

\def\beq{\begin{equation}}
\def\eeq{\end{equation}}
\def\eq{\end{equation}}
\def\to{\rightarrow}

\def\bsg{\ifmmode B\to X_s\gamma\else $B\to X_s\gamma$\fi}
\def\bsll{\ifmmode B\to X_s\ell^+\ell^-\else $B\to X_s\ell^+\ell^-$\fi}
\def\bstt{\ifmmode B\to X_s\tau^+\tau^-\else $B\to X_s\tau^+\tau^-$\fi}
\def\shat{\ifmmode \hat{s}\else $\hat{s}$\fi}

\newcommand{\newc}{\newcommand}

\newc{\lcal}{\int {\cal L}dt}

\newc{\LSP}{{\widetilde{\chi}^0_1}}
\newc{\stauR}{{\widetilde{\tau}_R}}
\newc{\stau}{{\widetilde{\tau}_1}}
\newc{\mstop}{m_{\widetilde{t}}}
\newc{\mHpm}{m_{H^\pm}}
\newc{\simgt}{\lower.7ex\hbox{$\;\stackrel{\textstyle>}{\sim}\;$}}
\newc{\simlt}{\lower.7ex\hbox{$\;\stackrel{\textstyle<}{\sim}\;$}}
\newc{\ie}{{\it i.e.}}
\newc{\etal}{{\it et al.}}
\newc{\eg}{{\it e.g.}}
\newc{\kev}{\hbox{\rm\,keV}}
\newc{\mev}{\hbox{\rm\,MeV}}
\newc{\gev}{\hbox{\rm\,GeV}}
\newc{\tev}{\hbox{\rm\,TeV}}
\newc{\xpb}{\hbox{\rm\, pb}}
\newc{\xfb}{\hbox{\rm\, fb}}

\newc{\mtop}{m_t}
\newc{\mbot}{m_b}
\newc{\mz}{m_Z}
\newc{\mw}{M_W}
\newc{\alphasmz}{\alpha_s(m_Z^2)}
\newc{\swsq}{\sin^2\theta_W}
\newc{\tw}{\tan\theta_W}
\newc{\cw}{\cos\theta_W}
\newc{\sw}{\sin\theta_W}
\newc{\BR}{\hbox{\rm BR}}
\newc{\zbb}{Z\to b\bar}
\newc{\Gb}{\Gamma (Z\to b\bar b)}
\newc{\Gh}{\Gamma (Z\to \hbox{\rm hadrons})}
\newc{\rbsm}{R_b^\hbox{\rm sm}}
\newc{\rbsusy}{R_b^\hbox{\rm susy}}
\newc{\drb}{\delta R_b}

\newc{\sgn}{\mbox{sgn}}
\newc{\tbeta}{\tan\beta}
\newc{\uL}{{\widetilde{u}_L}}
\newc{\uR}{{\widetilde{u}_R}}
\newc{\cL}{{\widetilde{c}_L}}
\newc{\cR}{{\widetilde{c}_R}}
\newc{\tL}{{\widetilde{t}_L}}
\newc{\tR}{{\widetilde{t}_R}}
\newc{\dL}{{\widetilde{d}_L}}
\newc{\dR}{{\widetilde{d}_R}}
\newc{\sL}{{\widetilde{s}_L}}
\newc{\sR}{{\widetilde{s}_R}}
\newc{\bL}{{\widetilde{b}_L}}
\newc{\bR}{{\widetilde{b}_R}}
\newc{\eL}{{\widetilde{e}_L}}
\newc{\eR}{{\widetilde{e}_R}}
\newc{\mhp}{m_{H^\pm}}
\newc{\mhalf}{m_{1/2}}
\newc{\emt}{{e/\mu /\tau}}

\newc{\lR}{\widetilde{l}_R}
\newc{\lL}{\widetilde{l}_L}
\newc{\nL}{\widetilde{\nu}_L}
\newc{\naa}{\widetilde{\chi}^0_1}
\newc{\nbb}{\widetilde{\chi}^0_2}
\newc{\ncc}{\widetilde{\chi}^0_3}
\newc{\ndd}{\widetilde{\chi}^0_4}
\newc{\nee}{\widetilde{\chi}^0_5}
\newc{\nff}{\widetilde{\chi}^0_6}
\newc{\caa}{\widetilde{\chi}^{\pm}_1}
\newc{\cbb}{\widetilde{\chi}^{\pm}_2}
\newc{\phit}{\phi_t}
\newc{\phib}{\phi_b}
\newc{\phiew}{\phi_{ew}}
\newc{\htz}{h^0_t}
\newc{\hbz}{h^0_b}
\newc{\hewz}{h^0_{ew}}
\newc{\hsmz}{h^0_{sm}}
\newc{\huz}{h^0_u}
\newc{\hsusyz}{h^0_{susy}}

\def\slashchar#1{\setbox0=\hbox{$#1$}           % set a box for #1
   \dimen0=\wd0                                 % and get its size
   \setbox1=\hbox{/} \dimen1=\wd1               % get size of /
   \ifdim\dimen0>\dimen1                        % #1 is bigger
      \rlap{\hbox to \dimen0{\hfil/\hfil}}      % so center / in box
      #1                                        % and print #1
   \else                                        % / is bigger
      \rlap{\hbox to \dimen1{\hfil$#1$\hfil}}   % so center #1
      /                                         % and print /
   \fi}                                         %
\catcode`@=11
\long\def\@caption#1[#2]#3{\par\addcontentsline{\csname
  ext@#1\endcsname}{#1}{\protect\numberline{\csname
  the#1\endcsname}{\ignorespaces #2}}\begingroup
    \small
    \@parboxrestore
    \@makecaption{\csname fnum@#1\endcsname}{\ignorespaces #3}\par
  \endgroup}
\catcode`@=12

\def\simlt{\stackrel{<}{{}_\sim}}
\def\simgt{\stackrel{>}{{}_\sim}}

\begin{document}

\baselineskip=18pt

\begin{titlepage}
\begin{flushright}
IZTECH-P-2009/04
\end{flushright}

\begin{center}
\vspace{1cm}

{\Large \bf General Tensor Lagrangians  from Gravitational
Higgs Mechanism}

\vspace{0.5cm}

{\bf Durmu{\c s} A. Demir$^{a}$ and N. K. Pak$^{b}$}

\vspace{.8cm}

{\it $^a$ Department of Physics, {\.I}zmir Institute of
Technology, TR35430
{\.I}zmir, Turkey\\
$^b$ Department of Physics, Middle East Technical University,
TR06530 Ankara, Turkey}

\end{center}
\vspace{1cm}

\begin{abstract}
\medskip
The gravitational Higgs mechanism proposed by 't Hooft in
\texttt{arXiv:0708.3184} involves the spacetime metric $g_{\mu
\nu}$ as well as the induced metric $\overline{g}_{\mu \nu}
\propto \eta_{a b} \partial_{\mu} \phi^a \partial_{\nu} \phi^b$
where $\phi^{a}$ $(a=0,\dots,3$), as we call it, break all four
diffeomorphisms spontaneously via the vacuum expectation values
$\langle \phi^a \rangle \propto x^a$. In this framework, we
construct and analyze the most general action density in terms
of various invariants involving the curvature tensors,
connexion coefficients, and the contractions and the
determinants of the two metric fields. We show that this action
admits a consistent expansion about the flat background such
that the resulting Lagrangian possesses several novel features
not found in the linearized Einstein-Hilbert Lagrangian with
Fierz-Pauli mass term (LEHL-FP): $(i)$ its kinetic part
generalizes that of LELHL-FP by weighing the corresponding
structures with certain coefficients generated by invariants,
$(ii)$ the entire Lagrangian is ghost-- and tachyon--free for
mass terms not necessarily in the Fierz-Pauli form, and,
$(iii)$ a consistent mass term is generated with no apparent
need to higher derivative couplings.



\end{abstract}

\bigskip
\bigskip

\begin{flushleft}
IZTECH-P-2009/04 \\
April 2009
\end{flushleft}

\end{titlepage}

%%%%%%%%%%%%%%%%%%%%%%%%%%%%%%%%%%%%%%%%%%%%%%%%%%%%%%%%%%%%%%%
\tableofcontents

\section{Introduction and Motivation}

In general, massive fields with spin $s\geq 1$ possess $2 s -1$
longitudinal components not found in their massless limit. These
extra components directly couple to the conserved currents, and their
effects do not necessarily disappear in the limit of vanishing
mass. Therefore, there arises a discontinuity in the field's mass, and
it renders the associated scattering amplitudes unphysical. This
phenomenon is known to occur in non-Abelian gauge theories
\cite{vainshtein} as well as (the linearized) gravity \cite{vDVZ}.
Indeed, the linearized Einstein-Hilbert action, linearized about the flat spacetime in
metric perturbations,
\begin{eqnarray}
\label{hmunu} h_{\mu \nu} \equiv g_{\mu \nu} - \eta_{\mu \nu}
\end{eqnarray}
admits a mass term of the form
\begin{eqnarray}
\label{mass} {\cal{L}}_{mass} = - \frac{1}{4} M_{Pl}^2 m_{g}^2
\left( h_{\alpha \beta} h^{\alpha \beta} - \zeta
h^{\alpha}_{\alpha} h^{\beta}_{\beta}\right)
\end{eqnarray}
where, $\zeta=1$ strictly, for the tensor theory to be ghost-free
\cite{fierz-pauli}. In other words, it is only and only for
$\zeta=1$, the trace mode $h \equiv h^{\alpha}_{\alpha}$,
which is a ghost as it possesses negative energy \cite{deser},
decouples from the rest. As mentioned before, this tensor theory
is discontinuous as $m_g \rightarrow 0$ \cite{vDVZ}.

A highly important feature of the Fierz-Pauli mass term
(\ref{mass}), worthy of emphasizing here, is that it holds only in
the linearized scheme. In other words, this very structure does
not admit any nonlinear completion obeying general covariance.
This immediately follows from the fact that, among the geometrical
quantities pertaining to the spacetime manifold, there is no
source, other than the determinant of the metric $
\mbox{det}\left(g_{\mu \nu}\right)$, for a non-derivative
structure like the Fierz-Pauli mass term. However, one immediately
runs into difficulties while trying to generate (\ref{mass}), if
$\mbox{det}\left(g_{\mu \nu}\right)$ is the only source. Indeed, a likely choice would be
to augment Einstein-Hilbert action by  vacuum energy
contribution, $\sqrt{- \mbox{det}\left(g_{\mu \nu}\right)} V_{vac}$. However, this energy component
cannot generate the graviton mass term correctly since, in the first
place, background geometry is wrong and, secondly, the quadratic part
of the linearized $\sqrt{-\mbox{det}\left(g_{\mu \nu}\right)}$ yields (\ref{mass}) with $\zeta=1/2$ not
$\zeta=1$. As a step further, one can imagine including higher
powers of $\sqrt{-\mbox{det}\left(g_{\mu \nu}\right)}$ for generating (\ref{mass}). However, this
is simply impossible in general relativity since $\mbox{det}\left(g_{\mu \nu}\right)$ is a
scalar density and the general covariance gets blatantly broken
unless some other scalar density (the determinant of some tensor field
different than the metric) is appropriately incorporated into the
action.

For a resolution of these problems (as reviewed  in detail in
\cite{review}), it is considered convenient to start with an analysis
of the mass discontinuity. This problem is overcome, in
non-Abelian gauge theories, via the Higgs mechanism through which
the gauge field develops requisite longitudinal component by
swallowing the Goldstone boson generated by the spontaneous
breakdown of the gauge symmetry. The system, as a whole,
consists not only of the gauge field but also of the scalars, so
that number of degrees of freedom remain unchanged as the state
changes from symmetric to broken phase, and vice versa. In
analogy with non-Abelian gauge theories, recently 't Hooft
\cite{thooft} (see also the previous work \cite{nima} and
references therein), followed by \cite{others1} and
\cite{others2}, proposed a similar mechanism for gravity in
which graviton acquires mass via the spontaneous breakdown of the
diffeomorphism invariance (see the earlier works
\cite{percacci}, \cite{chamseddine}, \cite{kirsch} and
\cite{leclerc} for variant approaches).

In essence, what 't Hooft suggests is to introduce `scalar
coordinates' $\phi^{a}\left(x\right)$ $(a=0, 1, 2, 3)$ which
are functions of the `vector coordinates' $x^{\mu}$
($\mu=0,1,2,3$) of the spacetime (in the spirit of manifold
structures \cite{compact} utilized for spacetime
compactification). Nonvanishing vacuum expectation values
(VEVs) of these scalars define a `preference gauge', more
precisely a `preference frame' in which diffeomorphism
invariance is spontaneously broken, whence graviton acquires a
nonvanishing mass \cite{thooft}. This mass generation process,
compared to gauge theories, is complicated by the fact that the
number of the scalars eaten is higher than the number of
longitudinal degrees of freedom that should be generated. In
other words, nonunitary degrees of freedom must be eliminated
to have a physically sensible massive graviton. This point, in
the framework of \cite{thooft,others2}, turns out to require
higher derivative couplings in the action, in order to provide
additional structures to eliminate the non-unitary modes. One
important aspect of the present work will be to show that,
elimination of the non-unitary modes does not necessarily
necessitate such higher derivative terms.

By imposing the invariance under the shifts \cite{thooft}
\begin{eqnarray}
\label{shift}
\phi^a \rightarrow \phi^a + c^a
\end{eqnarray}
it is automatically guaranteed that $(i)$ $\phi^a$ cannot have
non-derivative interactions such as a mass term, $(ii)$
$\phi^{a}$ can interact only gravitationally via their kinetic
terms, and finally $(iii)$ all the effects of scalar
coordinates can be encoded into the `induced metric'
\begin{eqnarray}
\label{gbar} \overline{g}_{\mu \nu} = \frac{1}{M^4} \eta_{a b}
\nabla_{\mu} \phi^a \nabla_{\nu} \phi^b
\end{eqnarray}
which is made dimensionless by rescaling its right-hand side by
$M^{4}$, $M$ being a mass scale related to the VEV of the
scalars $\phi^a$. At this point, for definiteness and clarity,
it proves useful to dwell on the meanings and implications of
the `metric fields' to be used throughout the text:
\begin{itemize}
\item The 'metric' fields $g_{\mu \nu}$ and
    $\overline{g}_{\mu \nu}$ are actually spin-2 tensor
    fields. In the technical sense, they are not `metric'
    fields. In fact, the former encodes the geometry
    (gravitational field) while the latter sets the
    background geometry with respect to which one studies
    the dynamics of spin-2 excitations $h_{\mu \nu}$ via
    (\ref{hmunu}). The true metric in this whole setup is
    the flat Minkowski metric $\eta_{\mu \nu}$. In spirit,
    the setup mimics that of bimetric gravity
    \cite{bimetric}.
\item The $\overline{g}_{\mu \nu}$, induced by the scalars
    $\phi^a$,  is a tensor field that plays a role similar
    to that of the Higgs field in the spontaneously broken gauge
    theories. It equals $\eta_{\mu \nu}$ when all the four
    diffeomorphisms are broken spontaneously, and this sets
    the background geometry.
\item The internal metric of the scalars is the flat
    Minkowski metric $\eta_{a b}$, not $\delta_{a b}$. In
    fact, structure of the induced metric in (\ref{gbar})
    parallels the decomposition $g_{\mu \nu} = \eta_{a b}
    e^{a}_{\mu} e^{b}_{\nu}$ so that the two metrics are
    related by the exchange of the vierbein $e^{a}_{\mu}$ and
    the gradient of the scalars $\nabla_{\mu} \phi^{a}$.
\end{itemize}
These observations reveal the distinctions among $g_{\mu \nu}$,
$\overline{g}_{\mu \nu}$ and $\eta_{\mu \nu}$, though they will
all be called `metric' in what follows. The $g_{\mu \nu}$ and
$\overline{g}_{\mu \nu}$ will be treated as two coexisting
metric fields, though the latter reduces to $\eta_{\mu \nu}$
upon spontaneous diffeomorphism breaking.

In this work, {\it we make use of the coexisting $\overline{g}_{\mu
\nu}$ and $g_{\mu \nu}$ fields to write down the most general
action density, and show that it admits a consistent expansion
about the flat background such that the resulting Lagrangian
owns several novel features not found in the linearized
Einstein-Hilbert Lagrangian with Fierz-Pauli mass term
(LEHL-FP): $(i)$ its kinetic part generalizes that of LELHL-FP
by weighing the corresponding structures with generic
coefficients formed by those of the invariants present in the
action, $(ii)$ the total Lagrangian qualifies to be ghost-- and
tachyon--free even for $\zeta \neq 1$ provided that the Lagrangian
parameters satisfy certain consistency relations, and finally,
$(iii)$ a consistent mass term arises with no apparent need to
higher derivative couplings.}

The rest of the work is organized as follows. In Sec. 2 below,
we construct the action density after determining exhaustively
the invariants made out of metric tensors, curvature tensors,
and connexion coefficients. Also in this section, we  derive
the linearized action, determine the conditions on model
parameters, and elaborate upon the generalized nature of the
action by comparing it with the LEHL-FP framework, in regard to
various extra structures not found in LEHL-FP setup. In Sec. 3
we summarize our main findings, and conclude.

\section{The Action}
For determining the most general action density, it proves useful
to first prepare an inventory of the invariants. The non-derivative
invariants are those constructed out of the metric fields,
$\overline{g}_{\mu \nu}$ and $g_{\mu \nu}$. In this class, there
naturally arise two fundamental invariants
\begin{eqnarray}
\texttt{K} = g_{\mu \nu} \overline{g}^{\mu \nu}\,,\;\;\;
\texttt{D} = \frac{\mbox{det}(\overline{g}_{\mu
\nu})}{\mbox{det}\left({g}_{\mu \nu}\right)}
\end{eqnarray}
where $\texttt{ K}$ is related to the kinetic term of the
$\phi^a$, and  the $\texttt{D}$ to the ratio of the
determinants of the two metrics. It is clear that $\texttt{K}$
necessarily embodies a ghost state, since the scalar fields
$\phi^a$ possess an indefinite metric ( $\eta_{a b}$ instead of
$\delta_{a b}$). The invariants $\texttt{K}$ and $\texttt{D}$
are also invariant under the shift transformation
(\ref{shift}). In fact, any function of them are also
shift-invariant, and hence, they contribute to the action
density via generic functions of the form
$V_1\left(\texttt{K}\right)$, $V_{2}\left(\texttt{D}\right)$
and $V_{3}\left(1/\texttt{D}\right)$. These functions, as they
stand, serve as shift-invariant `potentials' for  the metric
field.

The derivative invariants, that is, those invariants which involve
derivatives of the metric fields consist of a number of
structures constructed from the curvature tensors and the connexion
coefficients. Concerning the former, one readily finds two
invariants
\begin{eqnarray}
\label{curv-invs1} \texttt{R}_1 = \overline{g}^{\mu \nu} g^{\alpha
\beta} {\cal{R}}_{\mu \alpha \nu \beta}\,,\;\; \texttt{R}_2 =
\overline{g}^{\mu \nu} \overline{g}^{\alpha \beta} {\cal{R}}_{\mu
\alpha \nu \beta}
\end{eqnarray}
each of which dynamically differing from the usual Ricci scalar
${\cal{R}} \equiv g^{\mu \nu} g^{\alpha \beta} {\cal{R}}_{\mu
\alpha \nu \beta}$. Not surprisingly, these are not the only
curvature invariants since, being a metric tensor, the induced
metric $\overline{g}_{\mu \nu}$ itself generates novel
structures paralleling those generated by ${g}_{\mu \nu}$. To
this end, in the spirit of constructing Levi-Civita connexion
$\Gamma$ from the metric tensor $g_{\mu \nu}$, one can
construct a different connexion
\begin{eqnarray}
\overline{\Gamma}^{\lambda}_{\mu \nu}  = \frac{1}{2}
\widehat{\overline{g}}^{\lambda \rho} \left( \partial_{\mu}
\overline{g}_{\nu \rho} + \partial_{\nu} \overline{g}_{\rho \mu} -
\partial_{\rho} \overline{g}_{\mu \nu} \right)
\end{eqnarray}
based on $\overline{g}_{\mu \nu}$, assuming that it is
invertible. Here, $\widehat{\overline{g}}^{\lambda \rho}$ is
the matrix inverse of the induced metric
${\overline{g}}^{\lambda \rho}$, that is,
$\widehat{\overline{g}}_{\lambda \rho} {\overline{g}}^{\rho
\gamma} = \delta^{\gamma}_{\lambda}$. It is worth emphasizing
again, that $\widehat{\overline{g}}^{\lambda \rho} \neq
{\overline{g}}^{\lambda \rho} \equiv g^{\lambda \kappa}
\overline{g}_{\kappa \theta} g^{\theta \rho}$. Needless to say,
$\overline{\Gamma}^{\lambda}_{\mu \nu}$ is compatible with
$\overline{g}_{\mu \nu}$ in full analogy with the compatibility
of the $\Gamma^{\lambda}_{\mu \nu}$ with the $g_{\mu \nu}$.

As with the connexion $\Gamma^{\lambda}_{\mu \nu}$, the new
connexion $\overline{\Gamma}^{\lambda}_{\mu \nu}$ also generates
its Riemann tensor $\overline{{\cal R}}_{\mu \alpha \nu \beta}$
from which, similar to (\ref{curv-invs1}), one constructs the
curvature invariants
\begin{eqnarray}
\label{curv-invs2} \overline{\texttt{R}}_1 = {g}^{\mu \nu}
\overline{g}^{\alpha \beta} \overline{{\cal{R}}}_{\mu \alpha \nu
\beta}\,,\;\; \overline{\texttt{R}}_2 = {g}^{\mu \nu} {g}^{\alpha
\beta} \overline{{\cal{R}}}_{\mu \alpha \nu \beta}
\end{eqnarray}
in addition to the Ricci scalar $\overline{{\cal{R}}} \equiv
\overline{g}^{\mu \nu} \overline{g}^{\alpha \beta}
\overline{{\cal{R}}}_{\mu \alpha \nu \beta}$.

Apart from (\ref{curv-invs1}) and (\ref{curv-invs2}) generated by
the curvature tensors of $g_{\mu \nu}$ and $\overline{g}_{\mu \nu}$,
there exist extra invariants generated by the connexion
coefficients. Indeed, the difference
\begin{eqnarray}
\label{f-def}
{\cal{F}}^{\lambda}_{\mu \nu} = \Gamma^{\lambda}_{\mu \nu} -
\overline{\Gamma}^{\lambda}_{\mu \nu}
\end{eqnarray}
is a rank (1,2) tensor, and its contractions give rise to
additional derivative invariants independent of the curvature
tensors \cite{bimetric}. Obviously, all the invariants stemming
from this tensor field necessarily involve even occurrences of
${\cal{F}}^{\lambda}_{\mu \nu}$. In fact, up to the quadratic
order, possible invariants read as
\begin{eqnarray}
\texttt{C}_1 &=& g^{\mu \nu} {\cal{F}}^{\alpha}_{\alpha \mu}
{\cal{F}}^{\beta}_{\beta \nu}\,,\;\texttt{C}_2 = g^{\mu \nu}
{\cal{F}}^{\alpha}_{\beta \mu} {\cal{F}}^{\beta}_{\alpha
\nu}\nonumber\\ \texttt{C}_3 &=& g^{\mu \nu}
{\cal{F}}^{\alpha}_{\alpha \beta} {\cal{F}}^{\beta}_{\mu \nu}\,,\;
\texttt{C}_4 = g^{\mu \nu} g^{\alpha \beta} g_{\lambda \rho}
{\cal{F}}^{\lambda}_{\alpha \mu} {\cal{F}}^{\rho}_{\beta
\nu}\nonumber\\
\overline{\texttt{C}}_1 &=& \overline{g}^{\mu \nu}
{\cal{F}}^{\alpha}_{\alpha \mu} {\cal{F}}^{\beta}_{\beta
\nu}\,,\;\overline{\texttt{C}}_2 = \overline{g}^{\mu \nu}
{\cal{F}}^{\alpha}_{\beta \mu} {\cal{F}}^{\beta}_{\alpha
\nu}\nonumber\\ \overline{\texttt{C}}_3 &=& \overline{g}^{\mu \nu}
{\cal{F}}^{\alpha}_{\alpha \beta} {\cal{F}}^{\beta}_{\mu \nu}\,,\;
\overline{\texttt{C}}_4 = \overline{g}^{\mu \nu} g^{\alpha \beta}
g_{\lambda \rho} {\cal{F}}^{\lambda}_{\alpha \mu}
{\cal{F}}^{\rho}_{\beta \nu}\nonumber\\
\overline{\texttt{C}}_5 &=& \overline{g}^{\mu \nu}
\overline{g}^{\alpha \beta} g_{\lambda \rho}
{\cal{F}}^{\lambda}_{\alpha \mu} {\cal{F}}^{\rho}_{\beta \nu}\,,\;
\overline{\texttt{C}}_6 = \overline{g}^{\mu \nu}
\overline{g}^{\alpha \beta} \overline{g}_{\lambda \rho}
{\cal{F}}^{\lambda}_{\alpha \mu} {\cal{F}}^{\rho}_{\beta
\nu}
\end{eqnarray}
Note that in these invariants indices on ${\cal{F}}^{\lambda}_{\mu \nu}$
are kept as in (\ref{f-def}), with no further lowering or raising operations.

Having determined all possible invariants in the presence of two
metric fields, a general action integral can be written as
\begin{eqnarray}
\label{action} S_{G} &=& \frac{1}{2} M_{Pl}^2 \int d^4x\, \sqrt{- \mbox{det}\left(g_{\mu \nu}\right)}
\Bigg\{{\cal{R}} + \overline{a}\,\overline{\cal{R}} + a_1
\texttt{R}_1 + a_2 \texttt{R}_2 + \overline{a}_1
\overline{\texttt{R}}_1 + \overline{a}_2 \overline{\texttt{R}}_2\nonumber\\ &+&
\sum_{i=1}^{4} c_i \texttt{C}_i + \sum_{j=1}^6
\overline{c}_j \overline{\texttt{C}}_j + V_1(\texttt{K}) +
V_2(\texttt{D}) + V_3(1/\texttt{D}) \Bigg\}
\end{eqnarray}
wherein the derivative invariants with higher mass dimension
(such as ${\cal{R}}^n$, $\texttt{C}_i^2$, $\texttt{R}_i^n$ with
$n\geq 2$) are ignored. Therefore, $\overline{a}$, $a_i$,
$\overline{a}_i$, $c_i$ and $\overline{c}_i$ are all
dimensionless constants. In spite of this restricted structure
of the sector of derivative invariants, the sector of the
non-derivative invariants, represented by the `potentials'
$V_{1,2,3}$, is kept as general as possible to cope with
the constraints that can be faced with while inducing a
consistent graviton mass term.

The action density in (\ref{action}) incorporates, in the
geometrical sector, all possible invariants within the
aforementioned limits. Obviously, this action provides a more
general framework than those in the existing literature,
as the original proposal of 't Hooft \cite{thooft}, and its
refinements \cite{others1,others2} involve only ${\cal{R}}$ and
$V_1(\texttt{K})$ contributions. The extra structures, as will
be shown in the next section, give rise to novel features
in relation to structuring of the background geometry, canceling
the tadpoles, and killing the ghosts.

\section{Higgsing Gravity}
The general covariance guarantees that physical quantities are
independent of the choice of the coordinates. For instance,
invariance of the Einstein-Hilbert action under the
infinitesimal coordinate transformations ($\epsilon$ being
infinitesimal)
\begin{eqnarray}
\label{reparamet}
x^{\mu} \rightarrow x^{\mu} - \epsilon^{\mu}(x)
\end{eqnarray}
reflects itself in the conservation law expressed by the
contracted differential Bianchi identity. This coordinate
transformation or `gauge transformation' gives rise to the
diffeomorphisms
\begin{eqnarray}
\label{diffeo}
\delta g_{\mu \nu} &=& \nabla_{\mu} \epsilon_{\nu} + \nabla_{\nu}
\epsilon_{\mu}\nonumber\\
\delta \phi^a &=& \nabla_{\mu} \phi^a \epsilon^{\mu}
\end{eqnarray}
as dictated by variations of the metric $g_{\mu \nu}$, and the
scalars $\phi^a$ under general coordinate transformations. This
very reparametrization invariance is the fundamental gauge
symmetry of the action (\ref{action}) in that its status --
exact or broken -- determines whether or not there exist
massive excitations in the spectrum. Before turning our
attention to the broken symmetry case, which is the main aim of
this work, we first briefly discuss the case of exact
diffeomorphism invariance, corresponding to the massless
graviton, for completeness.

\subsection{Massless Graviton}
In massless phase, reparametrization invariance is exact. The vacuum configuration
\begin{eqnarray}
\label{massless-vac}
\langle g_{\mu \nu} \rangle = \eta_{\mu \nu}\,,\; \langle \phi^a \rangle = 0\,,
\end{eqnarray}
for which $\langle \overline{g}_{\mu \nu}\rangle = 0$ obviously, can
be sustained, as all the curvature invariants vanish trivially, if the potential functions
satisfy the constraint
\begin{eqnarray}
\label{vac-cond0}
V_1(0) + V_2(0) + V_3(\infty) = 0\,.
\end{eqnarray}
There is a subtlety involving the invariants $\texttt{C}_i$ and
$\overline{\texttt{C}}_i$. That is while $\Gamma^{\lambda}_{\mu
\nu} = 0$ trivially for strictly flat metric, the connexion
$\overline{\Gamma}^{\lambda}_{\mu \nu}$ of the induced metric
appears to have $\frac{0}{0}$ indeterminacy. Nonetheless, this
indeterminacy does not mean that
$\overline{\Gamma}^{\lambda}_{\mu \nu}$ diverges. In fact, with
an appropriate regularization procedure, for example, $\langle \phi^a
\rangle = \delta_1 x^a + \delta_2^b x_b x^a$ with $\delta_2^a
\ll \delta_1$ and $\delta^a_2 \rightarrow 0$, it can be
adjusted to vanish. Consequently, the invariants $\texttt{C}_i$
and $\overline{\texttt{C}}_i$ all vanish in the symmetric
vacuum (\ref{massless-vac}) since $\langle
{\cal{F}}^{\lambda}_{\mu \nu} \rangle = 0$ therein.

The excitations about the vacuum configuration (\ref{massless-vac}) involve two propagating
degrees of freedom pertaining to $h_{\mu \nu}$ and four associated with $\phi^a$ since all
the diffeomorhisms in (\ref{diffeo}) are exact. These six degrees of freedom constitute the
symmetric phase of the fields and the interactions encoded in (\ref{action}).

We have to reiterate that this regularization procedure affects
only the massless case. That it is, once
$\overline{\Gamma}^{\lambda}_{\mu \nu}$ is regulated to vanish
when  $\overline{g}_{\mu \nu} \rightarrow 0$, as long as the
constraint in (\ref{vac-cond0}) is satisfied, one realizes the
massless gravity limit smoothly. For the massive case, which is
at the focus of the present work, this subtlety is irrelevant,
however.

\subsection{Massive Graviton}
In the massive phase, reparametrization invariance is spontaneously broken. Indeed, the
vacuum configuration ($M$ being the mass scale appearing in (\ref{gbar}))
\begin{eqnarray}
\label{massive-vac}
\langle g_{\mu \nu} \rangle = \eta_{\mu \nu}\,,\; \langle \phi^a \rangle = M^2 x^a\,,
\end{eqnarray}
breaks the diffeomorphism invariance spontaneously and thus, as in gauge theories, defines a
`preference frame' such that all temporal and spatial diffeomorphisms are broken by imposing
\begin{eqnarray}
\label{fix} \delta \phi^a = 0
\end{eqnarray}
in (\ref{diffeo}). This gauge fixing forces all four scalar fields $\phi^a$ to remain stuck to their VEVs in
(\ref{massive-vac}), leaving behind no scalar fluctuations to propagate.
Expectedly, again in complete similarity to gauge theories, this gauge fixing procedure
automatically renders all 10 components of $h_{\mu \nu}$ physical.
However, a massive tensor field can have only 5 propagating modes,
and thus, the 5 extra components should be eliminated by the
dynamics encoded in (\ref{action}).

Under the gauge fixing (\ref{fix}), the building blocks of the
invariants in (\ref{action}) can be systematically expanded about
(\ref{massive-vac}) as follows:
\begin{enumerate}
\item The scalars $\phi^a$ are fixed to their VEVs: $\phi^a = \langle \phi^a
\rangle = M^2 x^a$.

\item The quantities involving the spacetime metric $g_{\mu \nu}$ we expanded up to
quadratic order as
\begin{eqnarray}
\label{quant2}
g_{\mu \nu} &=& \eta_{\mu \nu} + h_{\mu \nu}\,,\nonumber\\
g^{\mu \nu} &=& \eta^{\mu \nu} - h^{\mu \nu} + h^{\mu \alpha}
h^{\nu}_{\alpha} + {\cal{O}}\left(h^3\right)\,, \nonumber\\
- \mbox{det}\left(g_{\mu \nu}\right) &=& 1 + h +
\frac{1}{2} h^2 - \frac{1}{2} h^{\alpha \beta} h_{\alpha \beta} +
{\cal{O}}\left(h^3\right)\,,\nonumber\\
\Gamma^{\lambda}_{\mu \nu} &=& \frac{1}{2} \left( \eta^{\lambda
\rho} - h^{\lambda \rho}\right) \left(
\partial_{\mu} h_{\nu \rho} + \partial_{\nu} h_{\rho \mu} -
\partial_{\rho} h_{\mu \nu} \right) + {\cal{O}}\left(h^3\right)\,.
\end{eqnarray}

\item The quantities related to the induced metric are
    expanded as
\begin{eqnarray}
\label{quant2p}
\overline{g}_{\mu \nu} &=& \eta_{\mu \nu}\,,\nonumber\\
\overline{g}^{\mu \nu} &=& \eta^{\mu \nu} - 2 h^{\mu \nu} + 3
h^{\mu \alpha} h^{\nu}_{\alpha} +  {\cal{O}}\left(h^3\right)\,, \nonumber\\
- \mbox{det}\left(\overline{g}_{\mu \nu}\right) &=& 1\,, \nonumber\\
\overline{\Gamma}^{\lambda}_{\mu \nu} &=& 0\,.
\end{eqnarray}
\end{enumerate}
In addition, one has $\widehat{\overline{g}}_{\mu \nu} =
\eta_{\mu \nu}$, and since $\widehat{\overline{g}}^{\mu \nu}$
is the matrix inverse of ${\overline{g}}^{\mu \nu}$, it is
immediately found that $\widehat{\overline{g}}^{\mu \nu} =
\eta^{\mu \nu}$.

Having (\ref{fix}), (\ref{quant2}) and (\ref{quant2p}) at hand,
one can readily expand the action density in (\ref{action}) about
the vacuum configuration (\ref{massive-vac}) to obtain the $h_{\mu \nu}$ Lagrangian.
To begin with, one notes that $\overline{\cal{R}}$, $\overline{\texttt{R}}_1$,
$\overline{\texttt{R}}_2$ all vanish identically, as follows from
(\ref{quant2p}). The rest of the derivative invariants give rise
to the action density
\begin{eqnarray}
\label{kinetic} &-& \frac{M_{Pl}^2}{2} \Bigg[ \left(1+ a_1 + a_2
\right) \left( \partial_{\lambda}
\partial_{\rho} h^{\lambda
\rho} - \Box h \right)\nonumber\\
&+&\frac{\widehat{a}}{4}\,
\partial_{\lambda} h^{\alpha \beta} \partial^{\lambda} h_{\alpha
\beta} + \frac{\widehat{b}}{2}\, \partial_{\lambda} h^{\lambda
\alpha}
\partial^{\rho} h_{\rho \alpha} + \frac{\widehat{c}}{2}\,
\partial_{\lambda} h  \partial_{\rho} h^{\lambda \rho} + \frac{
\widehat{d}}{4}\, \partial_{\lambda} h \partial^{\lambda} h \Bigg]
\end{eqnarray}
whose first line, linear in $h_{\mu \nu}$, can obviously be
discarded away since it is a total divergence. The quadratic terms
in the second line form the kinetic part of the total $h_{\mu \nu}$
action. The hatted coefficients herein read in terms of the
original ones in (\ref{action}) as follows
\begin{eqnarray}
\label{param} \widehat{a} &=& 1 + 3 a_1 + 5 a_2 + c_2 +
\overline{c}_2 - 3 ( c_4
+ \overline{c}_4 + \overline{c}_5 + \overline{c}_6 ) \nonumber\\
\widehat{b} &=& -1 - 3 a_1 - 5 a_2 + c_4 + \overline{c}_4 +
\overline{c}_5 + \overline{c}_6 - c_2 - \overline{c}_2\nonumber\\
\widehat{c} &=& 1 + 2 a_1 + 3 a_2 - \frac{1}{2} ( c_3 +
\overline{c}_3)\nonumber\\
\widehat{d} &=& - 1 - a_1 - a_2 - c_1 - \overline{c}_1\,.
\end{eqnarray}
In each coefficient, the right-hand side starts with $\pm 1$ which
is what would be found within LEHL-FP formalism, and the additional
terms represent the deviations due to the curvature invariants occurring in
the presence of $\overline{g}_{\mu \nu}$.

The non-derivative part of the action originates from the
`potentials' $V_{1,2,3}$ in (\ref{action}). Expanding them up to
quadratic order by using (\ref{fix}), (\ref{quant2}) and
(\ref{quant2p}), the action density turns out to be
${M_{Pl}^2}/{2}$ times
\begin{eqnarray}
\label{pot-expand}
&& V_1(4) + V_2(1) + V_3(1)\nonumber\\
&+& \left[ V_3^{\prime}(1) - V_1^{\prime}(4) - V_2^{\prime}(1) +
\frac{1}{2}\left(V_1(4) + V_2(1) + V_3(1)\right)\right] h
\nonumber\\
&+& \Big[ \frac{1}{2} \left(  V_3^{\prime}(1) - V_1^{\prime}(4) -
V_2^{\prime}(1) \right) + \frac{1}{8}\left(V_1(4) + V_2(1) +
V_3(1)\right)\nonumber\\&+& \frac{1}{2} \left( V_2^{\prime}(1) +
V_3^{\prime}(1) + V_1^{\prime\prime}(4) + V_2^{\prime\prime}(1) +
V_3^{\prime\prime}(1)\right) \Big] h^2\nonumber\\
&+& \Big[- \frac{1}{4}\left(V_1(4) + V_2(1) + V_3(1)\right) -
\frac{1}{2}\left( V_3^{\prime}(1) - V_1^{\prime}(4) -
V_2^{\prime}(1)\right)\nonumber\\ &+& \frac{1}{2}
V_1^{\prime}(4)\Big] h^{\alpha \beta} h_{\alpha \beta}
\end{eqnarray}
where primes on $V_i$'s denote derivatives with respect to their
arguments. This action density is subject to certain consistency
conditions beyond (\ref{vac-cond0}) found in the symmetric phase.
First, for the entire procedure to be consistent, the
background geometry must be flat Minkowski, that is, the total
vacuum energy must vanish
\begin{eqnarray}
\label{vac-cond} V_1(4) + V_2(1) + V_3(1) = 0\,.
\end{eqnarray}
Next, in (\ref{pot-expand}) the terms linear in $h$ must also
vanish
\begin{eqnarray}
\label{tadpole-cond} V_3^{\prime}(1) - V_1^{\prime}(4) -
V_2^{\prime}(1) = 0
\end{eqnarray}
as otherwise classical field configuration gets destabilized (or
tadpoles are generated in quantized theory).

Combining the kinetic part in (\ref{kinetic}) with the remnant of
(\ref{pot-expand}) after imposing (\ref{vac-cond}) and
(\ref{tadpole-cond}), the total $h_{\mu \nu}$ action  takes the form
\begin{eqnarray}
\label{tot} S_{L} &=& - \frac{M_{Pl}^2}{2} \int d^4 x \Bigg[
\frac{\widehat{a}}{4}\, \partial_{\lambda} h^{\alpha \beta}
\partial^{\lambda} h_{\alpha \beta} + \frac{\widehat{b}}{2}\,
\partial_{\lambda} h^{\lambda \alpha}
\partial^{\rho} h_{\rho \alpha} + \frac{\widehat{c}}{2}\,
\partial_{\lambda} h  \partial_{\rho} h^{\lambda \rho}\nonumber\\ &+& \frac{
\widehat{d}}{4}\, \partial_{\lambda} h \partial^{\lambda}
h + \frac{1}{4} m_g^2 \left( h^{\alpha \beta} h_{\alpha \beta} -
\zeta h^2 \right) \Bigg]
\end{eqnarray}
where the mass term, taken to be precisely in the form of
(\ref{mass}), involves the `graviton mass'
\begin{eqnarray}
\label{grav-mass} m_g^2 &=& - 2 V_1^{\prime}(4)
\end{eqnarray}
as well as
\begin{eqnarray}
\label{grav-zeta} \zeta &=& - \frac{1}{V_1^{\prime}(4)}
\left(V_1^{\prime}(4) + 2 V_2^{\prime}(1) + V_1^{\prime\prime}(4)
+ V_2^{\prime\prime}(1) + V_3^{\prime\prime}(1)\right)
\end{eqnarray}
where uses have been made of the conditions (\ref{vac-cond}) and
(\ref{tadpole-cond}). These conditions, that the
vacuum energy and tadpoles must vanish, impose strong
fine-tuning constraints on the potentials $V_{i}$. Their
violations destabilize the background geometry and render the whole
procedure inconsistent. Apart from them, there arise additional constraints
stemming from the $h_{\mu \nu}$ dynamics itself, as to be determined below.

The action (\ref{tot}), in form, embodies the most general quadratic-level
Lagrangian for a symmetric tensor field. In fact, it is precisely
the generic tensor theory setup studied in \cite{pvn}, which provides a detailed
analysis of propagating modes and elimination of the ghosts and tachyons.
Nonetheless, for the present analysis, it proves particularly useful to focus on the equations of
motion themselves, especially for a clear view of the dynamics of the scalar ghost $h$. The equations of motion for $h_{\mu \nu}$, as originate from the
extremization of (\ref{tot}), read as
\begin{eqnarray}
\label{eomx}
&&\widehat{a}\, \Box h_{\mu \nu} + \widehat{b} \left(
\partial_{\mu}
\partial_{\rho} h^{\rho}_{\nu} + \partial_{\nu} \partial_{\rho}
h^{\rho}_{\mu} \right) + \widehat{c}\, \eta_{\mu \nu}
\partial_{\rho} \partial_{\lambda} h^{\rho \lambda} +
\widehat{c}\,
\partial_{\mu} \partial_{\nu} h + \widehat{d}\, \eta_{\mu \nu} \Box
h\nonumber\\ && -m_g^2 \left( h_{\mu \nu} - \zeta h \eta_{\mu
\nu}\right) = 0
\end{eqnarray}
which can be mapped into dynamical equations of lower spin components by
repeatedly applying contraction and divergence operations. This way, the
trace component $h$ is found to obey
\begin{eqnarray}
\label{h-eq}
b_2 \Box^2 h + b_1 m_g^2 \Box h + b_0 m_g^4 h = 0
\end{eqnarray}
wherein
\begin{eqnarray}
b_2 &=& - \widehat{a}^2 - 2 \widehat{a} ( \widehat{b} +
\widehat{c} + 2 \widehat{d}) - 6 \widehat{b} \widehat{d} + 3
\widehat{c}^2\,,\nonumber\\
b_1 &=& - 2 ( (1-2 \zeta) \widehat{a} + (1-3 \zeta) \widehat{b} +
\widehat{c} + 2 \widehat{d})\,,\nonumber\\
b_0 &=& 4 \zeta - 1\,,
\end{eqnarray}
which involve the model parameters in (\ref{param}) at the order
indicated by their subscripts.

The equations of motion for the vector component
$\partial_{\mu} h^{\mu \nu}$ is
\begin{eqnarray}
\label{v-eq}
&&\left[ \left(-\widehat{a} - \widehat{b}\right) \Box +
m_{g}^2\right] \partial_{\mu} h^{\mu \nu}\nonumber\\
&& + \frac{1}{2 \widehat{b} + 4 \widehat{c}}
\left[\left(\widehat{a}\left(\widehat{b}+\widehat{c}\right)-\widehat{b}\left(\widehat{c}
- 2 \widehat{d}\right)\right) \Box - \left(\widehat{c} + \left(1-2
\zeta\right)\right) m_g^2 \right] \partial^{\nu} h = 0
\end{eqnarray}
This is coupled to (\ref{h-eq}) via the gradient of $h$.

The equations of motion (\ref{h-eq}) and (\ref{v-eq}) reveal the unphysical degrees of
freedom contained in $h_{\mu \nu}$. Indeed, as the first point to note, the trace
field $h$ is clearly a ghost, and therefore, it should be prohibited to propagate. This
is accomplished by requiring
\begin{eqnarray}
\label{b2b1}
b_2 = 0\,,\; b_1 = 0
\end{eqnarray}
in (\ref{h-eq}). The main consequence of these conditions is
that $h$ is eliminated from the 10 total degrees of freedom in
$h_{\mu \nu}$ since, from (\ref{h-eq}), $h=0$ follows
unambiguously (One might alternatively consider taking $b_0 =
0$ as this also satisfies (\ref{h-eq}). However, this choice
does not eliminate $h$ from the spectrum; moreover, it gives
rise to a ghosty graviton as it enforces $\zeta = 1/4$.).

%For the linearized Einstein-Hilbert action, $b_2$ vanishes
%automatically, and $b_1 \propto (1-\zeta)$. It is for this reason that killing the
%ghost in Fierz-Pauli action requires $\zeta = 1$ \cite{fierz-pauli}. For the generalized setup
%(\ref{tot}) this relation will gener

Setting $h=0$ in (\ref{v-eq}) reveals that the vector
$\partial_{\mu} h^{\mu \nu}$ is also a ghost, and its elimination
from the spectrum requires
\begin{eqnarray}
\label{a+b}
\widehat{a} + \widehat{b} = 0
\end{eqnarray}
as a further condition for having a ghost-free tensor theory.

Consequently, the equations of motion (\ref{eomx}), after eliminating scalar
ghost $h$ and vector ghost $\partial_{\mu} h^{\mu \nu}$, take the form
\begin{eqnarray}
\label{eqns}
h &=& 0\,,\nonumber\\
\partial_{\mu} h^{\mu \nu} &=& 0\,,\nonumber\\
\left(\widehat{a} \Box  - m_{g}^2\right) h_{\mu \nu}  &=& 0\,,
\end{eqnarray}
where, obviously, it is imperative to have
\begin{eqnarray}
\label{mg}
\widehat{a}>0\,, \; m_{g}^2 >0
\end{eqnarray}
for equations of motion (\ref{eqns}) to describe a non-ghost,
non-tachyonic, massive, free, spin-2 field. The wave equation
for $h_{\mu \nu}$ can be put into conventional form by
rescaling the action (\ref{tot}) by $1/\widehat{a}$. Positivity
of $m_g^2$, through (\ref{grav-mass}), implies that
$V_1^{\prime}(4) < 0$. The main implication of this, recalling
that $\texttt{K} = g_{\mu \nu} \overline{g}^{\mu \nu} = 4 - h +
{\cal{O}}\left(h^2\right)$, is that $V_1(\texttt{K})$ must
obtain negative slope around $\texttt{K} =4$ so as to invert
the signature of $\overline{g}_{\mu \nu}$ controlled by
$\eta_{a b}$. This switch of signature makes $m_g^2$ positive,
or equivalently, the graviton non-tachyonic.

The equations of motion (\ref{eqns}) hold only for free $h_{\mu
\nu}$. The matter sector can be incorporated into
geometrodynamics by augmenting the action (\ref{action}) with
\begin{eqnarray}
\Delta S_{G} = \int d^4 x \, \sqrt{- \mbox{det}\left(g_{\mu \nu}\right)} {\cal{L}}_{matter}\left(g, \overline{g}, \psi\right)
\end{eqnarray}
where $\psi$ stands for matter fields, collectively. This
add-on interaction causes the $h_{\mu \nu}$ action (\ref{tot})
to be extended by $-(1/2) h^{\mu \nu} T^{matter}_{\mu \nu}$.
Consequently, unless the matter stress tensor $T^{matter}_{\mu
\nu}$ possesses certain special features, all components of
$h_{\mu \nu}$, excluding the vector ghost $\partial_{\mu}
h^{\mu \nu}$, couple to and affected by $T^{matter}_{\mu \nu}$.
This implies, in particular, that the equation of motion of $h$
(\ref{h-eq}) possesses an inhomogeneity involving the trace of
$T^{matter}_{\mu \nu}$. If the matter sector is not
conformal invariant, which indeed is not, it becomes impossible
to eliminate $h$, or equivalently, to obtain $h=0$. This
problem was already noticed by 't Hooft in \cite{thooft}, and a
resolution was suggested: Similar to the potentials $V_{2,3}$,
the matter Lagrangian should also depend on the metric tensor
via the determinantal invariant $\texttt{D}$. More explicitly,
the matter Lagrangian must have the specific structure
\begin{eqnarray}
{\cal{L}}_{matter}\left(g_{\mu \nu}, \overline{g}_{\mu \nu}, \psi\right) \equiv {\cal{L}}_{matter}\left(\texttt{D}^{1/6} g_{\mu \nu}, \psi\right)
\end{eqnarray}
so that the scalar ghost $h$ gets eliminated despite the presence of matter.

Having reached a physically sensible picture of massive
graviton, at this stage it could be useful to perform a global
analysis of the resulting constraints on the model parameters.
Tabulated in Table 1 are the constraints imposed by having
ghost-- and tachyon--free massive gravity. The implications or
status of the constraints are shown for both the LEHL-FP and the
present model. In the linearized Einstein-Hilbert action with
Fierz-Pauli type mass term (the LEHL-FP framework), the
parameters $\widehat{a},\dots, \widehat{d}$ take on rather
specific values such that all the bounds and constraints are
satisfied trivially (designated by the symbol $\surd$ in the
third column). The only exception is $\zeta$, namely $b_1=0$
requires $\zeta =1$ which is the unique value of $\zeta$
\cite{pvn,others2} for the Fierz-Pauli mass term defined in
(\ref{mass}).

Concerning the model under investigation, constraints and
resulting bounds on or relations among the model parameters are
displayed in the fourth column of Table 1. The fact that
$\widehat{a},\dots,\widehat{d}$ deviate from LEHL-FP limit due
to nonvanishing $a_i$, $c_i$ and $\overline{c}_i$
contributions, leaves important impact on parametric relations
arising in response to bounds and constraints. In particular,
none of the constraints (listed in the second column) is
satisfied trivially; each is realized at the expense of
imposing a further relation, which itself leads to the
determination or bounding of a certain parameter in terms of
the others. The constraints are not sufficient in number for a
full determination of the model parameters. Nevertheless,
various relations in the fourth column reflect the generalized
nature of (\ref{tot}) with respect to the LEHL-FP framework. A
highly important feature is that $\zeta$ is forced to have a
specific relation to  $\widehat{a}$, $\widehat{b}$ and
$\widehat{c}$. However, as a direct consequence of the
gravitational Higgs mechanism, the same parameter is related
also to the potential functions $V_{i}$, at specific values of
these arguments as depicted in (\ref{grav-zeta}). Therefore,
elimination of the scalar ghost imposes a direct correlation
between the derivative and non-derivative sectors in
(\ref{action}) by forcing $\zeta$ to be equal to
\begin{eqnarray}
\label{2step}
\zeta = \frac{\widehat{a}^2 - 3 \widehat{a} \widehat{c} + 3 \widehat{c}^2}{\widehat{a}^2} &=& \frac{1}{4 ( 1+ 3 {a_1}+5 {a_2}+ c_2 + \overline{c}_2)^2} \Bigg[
4 + 12 {a_1}^2 + 28 a_2^2 \nonumber\\ &+& 6 a_1 (2+ 6 {a_2}- c_3 - \overline{c}_3) + 4 (c_2 + \overline{c}_2) (c_2 + \overline{c}_2 -1) \nonumber\\ &+&
2 {a_2} (8 + 2 (c_2 + \overline{c}_2) - 3 (c_3 + \overline{c}_3))\nonumber\\ &+& 3 (c_3 + \overline{c}_3)
   (- 2 + c_3 + \overline{c}_3 + 2(c_2 + \overline{c}_2)) \Bigg]\nonumber\\
   &=& - \frac{1}{V_1^{\prime}(4)}
\left(V_1^{\prime}(4) + 2 V_2^{\prime}(1) + V_1^{\prime\prime}(4)
+ V_2^{\prime\prime}(1) + V_3^{\prime\prime}(1)\right)
\end{eqnarray}
where use has been made of (\ref{param}) and (\ref{a+b}) in the
second step. This equality can be used to eliminate one of the
unknowns. For instance, it can be used to solve $c_2 +
\overline{c}_2$ in terms of $c_3 +\overline{c}_3$, $a_1$, $a_2$
and the potentials in the second line. The solution, after
replacing in the third row of Table 1, determines $c_1
+\overline{c}_1$ in terms of $c_3 +\overline{c}_3$, $a_1$,
$a_2$ and the potentials. This, however, does not bring any
important novelty in that $c_1 + \overline{c}_1$ just gets
expressed in terms of the potential functions instead of $c_2 +
\overline{c}_2$. Nonetheless, extraction of $c_2 +
\overline{c}_2$ from (\ref{2step}) gives some useful bounds in
light of the constraint $\widehat{a} > 0$ (implying $1+ 3 a_1 +
5 a_2 + c_2 + \overline{c}_2> 0$ as shown in the fourth row of
Table 1). Indeed, one finds that
\begin{eqnarray}
\label{13a15a2}
1+ 3 a_1 + 5 a_2 + c_2 + \overline{c}_2 = \frac{3}{4 \varpi + 2} \left( 2 + 4 a_1 + 6 a_2 - c_3 -\overline{c}_3\right)
\left[1\pm \sqrt{- \frac{1}{3}(4 \varpi + 5)}\right]
\end{eqnarray}
where
\begin{eqnarray}
-1-\varpi \equiv - \frac{1}{V_1^{\prime}(4)}
\left(V_1^{\prime}(4) + 2 V_2^{\prime}(1) + V_1^{\prime\prime}(4)
+ V_2^{\prime\prime}(1) + V_3^{\prime\prime}(1)\right)
\end{eqnarray}
which equals the second line of (\ref{2step}). The $\pm$ signs
correspond to the two solutions of $c_2 + \overline{c}_2$ as
extracted from (\ref{2step}). This quantity can be guaranteed
to be positive by various combinations of signs and magnitudes
of the parameters at the right-hand side. On the other hand,
the parameter $\varpi$ is bounded by
\begin{eqnarray}
\varpi < - \frac{5}{4}
\end{eqnarray}
as follows from the terms in the radical sign in
(\ref{13a15a2}). This then gives rise to the constraint
\begin{eqnarray}
\label{new-pot-bound}
2 V_2^{\prime}(1) + V_1^{\prime\prime}(4) + V_2^{\prime\prime}(1) + V_3^{\prime\prime}(1) > \frac{5}{4} \left|V_1^{\prime}(4)\right|
\end{eqnarray}
after using the inequality $V_1^{\prime}(4) < 0$ for graviton
to be non-tachyonic (as indicated in the fifth row of Table 1).
However, there is more than this. Indeed, after using
(\ref{new-pot-bound}) in the definition of $\zeta$ in
(\ref{grav-zeta}), one arrives at the bound
\begin{eqnarray}
\label{zeta-bound}
\zeta > \frac{1}{4}
\end{eqnarray}
which clearly shows that $\zeta$ is positive yet does not need
to take its value preferred by the Fierz-Pauli mass term. This
bound is indicated in the last row of Table 1.

It is clear that the elimination of the scalar ghost does only
put a bound on $\zeta$ as given in (\ref{zeta-bound}). For
instance, there is no obligation to have one or all of the
$V_i^{\prime \prime}$ to be nonzero. Indeed, they can all
vanish without causing a problem, provided that
$V_{2}^{\prime}(1)$ assumes an appropriate value to satisfy
(\ref{new-pot-bound}). In this sense, thanks to the inclusion
of determinantal invariants, $V_2\left(\texttt{D}\right)$ and
$V_3\left(1/\texttt{D}\right)$ in (\ref{action}), it becomes
possible to induce a physically consistent graviton mass with
no fundamental need to the higher derivative couplings. This is
a novel feature not found in \cite{others2}, wherein it is
shown that the existence of higher derivative couplings are
essential for eliminating the $h$.

\begin{table}[t]
\begin{center}
\begin{tabular}{|l|l|l|l|}
  \hline
Equation & Relation &$\begin{array}{c} \mbox{LEHL-FP Model}\\ (\widehat{a} = -\widehat{b} = \widehat{c} = -\widehat{d} = 1)\end{array}$
        & $\begin{array}{c} \mbox{Present Model [Eq.(\ref{action}) or Eq.(\ref{tot})]}\\ (\mbox{see Eq.(\ref{param}) for parameters}) \end{array}$
         \\ \hline\hline
Eq. (\ref{a+b}) &  $\widehat{a} + \widehat{b} = 0$ & $\surd\;$ &  $\surd\;$ if $\;\; c_4 + \overline{c}_4 + \overline{c}_5 + \overline{c}_6 = 0$ \\ \hline
Eq. (\ref{b2b1}) &  $b_2 = 0$ & $\surd\;$ & $\begin{array}{l} \surd\; \mbox{if} \;\; \widehat{d} = - \left((\widehat{a}-\widehat{c})^2 +
2 \widehat{c}^2\right)/2 \widehat{a}\\ \left( \begin{array}{l} c_{1} + \overline{c}_1 =  \frac{1}{8 (1+ 3 {a_1}+5
   {a_2}+ {c_2 + \overline{c}_2})} \Bigg[ 3 (c_3 + \overline{c}_3)^2\\ - 4 (c_3 + \overline{c}_3) ( 2+ 3 {a_1}+
   4{a_2}- c_2 - \overline{c}_2 )\\ + 4 \Big(3 a_1^2+12 a_1 a_2 + 12 a_2^2 + (c_2 + \overline{c}_2)^2
   \\+ 2 (a_2 -1) (c_2 + \overline{c}_2)\Big)\Bigg]\end{array}
   \right) \end{array}$ \\ \hline
Eq. (\ref{mg}) & $\widehat{a} > 0$ & $\surd$ & $\surd\;$ if
$\;\; 1+ 3 a_1 + 5 a_2 + c_2 + \overline{c}_2 > 0$  \\ \hline
Eq. (\ref{mg}) & $m_g^2 > 0$ & $\surd$ & $\surd\;$ if $\;\;
V_1^{\prime}(4) < 0$ \\ \hline
Eq. (\ref{b2b1}) & $b_1 = 0$ & $\surd\;$ if $\zeta = 1$ & $\surd\;$
if $\;\; \zeta = \left(\widehat{a}^2 - 3 \widehat{a}
\widehat{c} + 3 \widehat{c}^2\right)/\widehat{a}^2 > \frac{1}{4}$ \\ \hline
\end{tabular}
\vspace{0.5cm}
\end{center}
\label{table1}
\caption{Constraints on the model parameters for having a ghost-- and tachyon--free massive graviton. The implication or status of
each constraint is depicted for LEHL-FP (the third column) and the present model (the fourth column). The symbol $\surd$ means that
a given constraint is satisfied trivially (as happens for LEHL-FP for all constraints except for $b_1 = 0$) or upon the imposition of
a condition which itself constrains or determines certain parameters in terms of the others (as happens for the present formalism
in all cases).}
\end{table}

\section{Conclusion}
In this work, by exploiting the coexistence of two metric
fields $\overline{g}_{\mu \nu}$ and $g_{\mu \nu}$ in the
gravitational Higgs mechanism proposed by 't Hooft
\cite{thooft}, we have constructed and studied the most general
action functional (\ref{action}). The action involves both
derivative (originating from the curvature tensors and the connexion
coefficients) as well as non-derivative (originating from both
$g_{\mu \nu}$ and $\overline{g}_{\mu \nu}$) invariants.

We have shown that the action density in (\ref{tot}) admits a
consistent expansion about the flat background such that the
resulting Lagrangian (\ref{tot}) possesses several novel
features not found in the linearized Einstein-Hilbert
Lagrangian with the Fierz-Pauli mass term. First of all, its
kinetic part generalizes that of the LELHL-FP framework by
weighing the corresponding structures with generic coefficients
(\ref{param}) generated by the invariants present in
(\ref{action}). Next, a ghost-- and tachyon--free massive
gravity theory arises, once the conditions in the Table 1 are
met. In particular, the absence of the ghosts and the tachyons
does not require $\zeta =1$; it takes a general value shown in
the fifth row of the Table 1, provided that the constraint
(\ref{2step}) is respected.

It is true that the action in (11) contains various independent
structures which come with independent coefficients. We have
checked that one can eliminate several of these by making use
of the relations stemming from the constraints tabled in the
fourth column of the Table 1. However, as the number of the
constraints is fewer than the number of parameters, there are
yet several free parameters left over in the scheme after
eliminating as many of these as the constraints enable us to
do. The parameters $c_3 + \overline{c}_3$, $a_1$, $a_2$ and
various potential functions remain as essentially free
parameters (as long as (\ref{new-pot-bound}) and the bounds in
fourth and fifth rows of Table 1 are satisfied).

Another important feature concerns the nature of the non-derivative
invariants. The inclusion of the determinantal invariants
facilitates generation of the graviton mass term with no apparent
need to the higher derivative couplings. In other words, the
potentials $V_i$ can have vanishing derivatives at the second
and the higher orders, yet a physically meaningful graviton mass
still arises, as shown in (\ref{2step}).

In the entire text the focus of our attention was on the massive gravity, only. However,
this does not need to be so. Indeed, the action (\ref{tot}) does also
describe of glueball dynamics in QCD after the replacements $M_{Pl} \rightarrow \Lambda_{QCD}$
and $m_{g} \sim 1 {\rm GeV}$. Therefore, generality of (\ref{tot}) can also
provide useful tools for exploring the glueballs in QCD.

\section{Acknowledgements}
The work of D. D. was supported by  the Alexander von
Humboldt-Stiftung Friedrich Wilhelm Bessel-Forschungspreise and
by the Turkish Academy of Sciences via GEBIP grant. D. D.
thanks for hospitality to the Theory Group at DESY, Hamburg
where this work was started. The research of N.K.P. was
supported by Turkish Academy of Sciences through a membership
research grant.


\begin{thebibliography}{999}

\bibitem{vainshtein}
A.~I.~Vainshtein and I.~B.~Khriplovich,
  %``On the zero-mass limit and renormalizability in the theory of massive
  %yang-mills field,''
  Sov.\ J.\ Nucl.\ Phys.\  {\bf 13} (1971) 111;
  %%CITATION = YAFIA,13,198;%%
A.~Vainshtein,
  %``Massive Gravity,''
  Surveys High Energ.\ Phys.\  {\bf 20} (2006) 5.
  %%CITATION = SHEPD,20,5;%%

\bibitem{vDVZ}
Y.~Iwasaki,
  %``Consistency condition for propagators,''
  Phys.\ Rev.\  D {\bf 2}, 2255 (1970);
  %%CITATION = PHRVA,D2,2255;%%
H.~van Dam and M.~J.~G.~Veltman,
  %``Massive And Massless Yang-Mills And Gravitational Fields,''
  Nucl.\ Phys.\  B {\bf 22} (1970) 397;
  %%CITATION = NUPHA,B22,397;%%
V.~I.~Zakharov,
  %``Linearized gravitation theory and the graviton mass,''
  JETP Lett.\  {\bf 12} (1970) 312
  [Pisma Zh.\ Eksp.\ Teor.\ Fiz.\  {\bf 12} (1970) 447].
  %%CITATION = ZFPRA,12,447;%%

\bibitem{fierz-pauli}
M.~Fierz and W.~Pauli,
  %``On relativistic wave equations for particles of arbitrary spin in an
  %electromagnetic field,''
  Proc.\ Roy.\ Soc.\ Lond.\  A {\bf 173} (1939) 211.
  %%CITATION = PRSLA,A173,211;%%

\bibitem{deser}
D.~G.~Boulware and S.~Deser,
  %``Can gravitation have a finite range?,''
  Phys.\ Rev.\  D {\bf 6} (1972) 3368.
  %%CITATION = PHRVA,D6,3368;%%

\bibitem{review} V.~A.~Rubakov and P.~G.~Tinyakov,
  %``Infrared-modified gravities and massive gravitons,''
  Phys.\ Usp.\  {\bf 51} (2008) 759
  [arXiv:0802.4379 [hep-th]].
  %%CITATION = PHUSE,51,759;%%


\bibitem{thooft}
G.~'t Hooft,
  %``Unitarity in the Brout-Englert-Higgs Mechanism for Gravity,''
  arXiv:0708.3184 [hep-th].
  %%CITATION = ARXIV:0708.3184;%%

\bibitem{nima} N.~Arkani-Hamed, H.~Georgi and M.~D.~Schwartz,
  %``Effective field theory for massive gravitons and gravity in theory space,''
  Annals Phys.\  {\bf 305}, 96 (2003)
  [arXiv:hep-th/0210184].
  %%CITATION = APNYA,305,96;%%



\bibitem{others1}
Z.~Kakushadze,
  %``Gravitational Higgs Mechanism and Massive Gravity,''
  Int.\ J.\ Mod.\ Phys.\  A {\bf 23} (2008) 1581
  [arXiv:0709.1673 [hep-th]];
  %%CITATION = IMPAE,A23,1581;%%
I.~Oda,
  %``Gravitational Higgs Mechanism with a Topological Term,''
  arXiv:0709.2419 [hep-th].
  %%CITATION = ARXIV:0709.2419;%%

\bibitem{others2}
Z.~Kakushadze,
  %``Massive Gravity in Minkowski Space via Gravitational Higgs Mechanism,''
  Phys.\ Rev.\  D {\bf 77} (2008) 024001
  [arXiv:0710.1061 [hep-th]].
  %%CITATION = PHRVA,D77,024001;%%

\bibitem{percacci}
R.~Percacci,
  %``The Higgs Phenomenon in Quantum Gravity,''
  Nucl.\ Phys.\  B {\bf 353} (1991) 271
  [arXiv:0712.3545 [hep-th]].
  %%CITATION = NUPHA,B353,271;%%

\bibitem{chamseddine}
A.~H.~Chamseddine,
  %``Spontaneous symmetry breaking for massive spin-2 interacting with
  %gravity,''
  Phys.\ Lett.\  B {\bf 557} (2003) 247
  [arXiv:hep-th/0301014].
  %%CITATION = PHLTA,B557,247;%%

\bibitem{kirsch}
I.~Kirsch,
  %``A Higgs mechanism for gravity,''
  Phys.\ Rev.\  D {\bf 72} (2005) 024001
  [arXiv:hep-th/0503024];
  %%CITATION = PHRVA,D72,024001;%%
N.~Boulanger and I.~Kirsch,
  %``A Higgs mechanism for gravity. II: Higher spin connections,''
  Phys.\ Rev.\  D {\bf 73}, 124023 (2006)
  [arXiv:hep-th/0602225].
  %%CITATION = PHRVA,D73,124023;%%


\bibitem{leclerc}
R.~Parthasarathy,
  %``ON SPACE-TIME COMPACTIFICATION INDUCED BY A GENERAL NONLINEAR SIGMA
  %MODEL,''
  Phys.\ Lett.\  B {\bf 181} (1986) 91;
  %%CITATION = PHLTA,B181,91;%%
M.~Leclerc,
  %``The Higgs sector of gravitational gauge theories,''
  Annals Phys.\  {\bf 321} (2006) 708
  [arXiv:gr-qc/0502005].
  %%CITATION = APNYA,321,708;%%

\bibitem{compact}
C.~Omero and R.~Percacci,
  %``Generalized Nonlinear Sigma Models In Curved Space And Spontaneous
  %Compactification,''
  Nucl.\ Phys.\  B {\bf 165} (1980) 351;
  %%CITATION = NUPHA,B165,351;%%
M.~Gell-Mann and B.~Zwiebach,
  %``Space-Time Compactification Due To Scalars,''
  Phys.\ Lett.\  B {\bf 141} (1984) 333;
  %%CITATION = PHLTA,B141,333;%%
D.~A.~Demir and B.~Pulice,
  %``Non-gravitating scalars and spacetime compactification,''
  Phys.\ Lett.\  B {\bf 638} (2006) 1
  [arXiv:hep-th/0605071].
  %%CITATION = PHLTA,B638,1;%%



\bibitem{pvn}
P.~Van Nieuwenhuizen,
  %``On Ghost-Free Tensor Lagrangians And Linearized Gravitation,''
  Nucl.\ Phys.\  B {\bf 60} (1973) 478.
  %%CITATION = NUPHA,B60,478;%%

\bibitem{bimetric} N.~Rosen,
  %``General Relativity and Flat Space. I,''
  Phys.\ Rev.\  {\bf 57}, 147 (1940);
  %%CITATION = PHRVA,57,147;%%
%``General Relativity and Flat Space. II,''
  Phys.\ Rev.\  {\bf 57}, 150 (1940);
  %%CITATION = PHRVA,57,150;%%
J.~W.~Moffat,
  %``Bimetric gravity theory, varying speed of light and the dimming of
  %supernovae,''
  Int.\ J.\ Mod.\ Phys.\  D {\bf 12}, 281 (2003)
  [arXiv:gr-qc/0202012].
  %%CITATION = IMPAE,D12,281;%%






\end{thebibliography}
\end{document}